\documentclass[twocolumn,showpacs,amsmath,amssymb,prd]{revtex4}

\usepackage{graphicx}
\usepackage{graphicx}
\usepackage{dcolumn}
\usepackage{bm}
\usepackage{epsf}

\def \agt{\gtrsim}
\def \alt{\lesssim}

\newcommand{\beq}{\begin{equation}}
\newcommand{\eeq}{\end{equation}}
\newcommand{\beqn}{\begin{eqnarray}}
\newcommand{\eeqn}{\end{eqnarray}}

\begin{document}

\title{Collapse of magnetized hypermassive neutron stars in general 
relativity}

\author{Matthew D. Duez$^1$}
\altaffiliation{Currently address:  Center for Radiophysics and Space Research,
Cornell University, Ithaca, NY 14853}

\author{Yuk Tung Liu$^1$}

\author{Stuart L. Shapiro$^1$}
\altaffiliation{Also at the Department of Astronomy and NCSA, University
of Illinois at Urbana-Champaign, Urbana, IL 61801}

\author{Masaru Shibata$^2$}

\author{Branson C. Stephens$^1$}
\affiliation{$^1$Department of Physics, University of Illinois at 
Urbana-Champaign, Urbana, IL 61801, USA \\
$^2$Graduate School of Arts and Sciences, 
University of Tokyo, Komaba, Meguro, Tokyo 153-8902, Japan}

\begin{abstract}
Hypermassive neutron stars (HMNSs) -- equilibrium 
configurations supported against 
collapse by rapid differential rotation -- are possible transient 
remnants of binary neutron star mergers. Using newly developed 
codes for magnetohydrodynamic simulations in dynamical spacetimes, 
we are able to track the evolution of a magnetized HMNS in full general 
relativity for the first time. We find that secular angular momentum 
transport due to magnetic braking and the magnetorotational instability 
results in the collapse of an HMNS to a rotating 
black hole, accompanied by a gravitational wave burst.  
The nascent black hole is 
surrounded by a hot, massive torus undergoing quasistationary accretion 
and a collimated magnetic field.  This scenario suggests 
that HMNS collapse is a possible 
candidate for the central engine of short gamma-ray bursts.
\end{abstract}
\pacs{04.25.Dm, 04.30.-w, 04.40.Dg}

\maketitle

Hypermassive neutron stars (HMNSs) figure prominently in several
relativistic astrophysical systems of current interest.  
Mass limits for nonrotating stars [the Oppenheimer-Volkoff
(OV) limit] and for rigidly rotating stars (the supramassive limit,
which is only about 20\% larger) can be significantly exceeded by
the presence of {\em differential} rotation~\cite{BSS}.  Stars with
masses greater than the supramassive limit are called hypermassive stars.
Mergers of binary neutron stars (BNSs) can lead to the 
formation of HMNS remnants.  This possibility was explored in 
Newtonian and post-Newtonian simulations~\cite{RS99,FR00}, and in 
full general relativity~\cite{STUa}. 
The latest relativistic BNS merger simulations with realistic equations of 
state (EOS)~\cite{STUb} 
confirm that HMNS formation is indeed a possible outcome. 
HMNSs can also result from core collapse of rotating massive stars, 
since rapid differential rotation can develop during the 
collapse.

Differentially rotating stars approach rigid rotation 
via transport of angular momentum on secular time scales. HMNSs,
however, cannot settle down to rigidly rotating equilibria since 
their masses exceed the maximum allowed by uniform rotation. Thus, 
`delayed' collapse to a black hole, and possibly mass loss, 
will follow transport of angular momentum
from the inner to the outer regions. 
Previous calculations
of HMNS collapse have focused on viscous angular momentum transport~\cite{DLSS}
and angular momentum loss due to gravitational radiation~\cite{STUb}.
In this Letter, we demonstrate black hole formation induced 
by seed magnetic fields in HMNSs. 

In any highly conducting astrophysical plasma, a frozen-in magnetic field 
can be amplified appreciably by gas compression or shear (e.g. differential
rotation).  Even when an initial seed magnetic field is weak, the
field can grow to influence significantly the system dynamics. 
There are at least two distinct effects
which amplify the magnetic field in a HMNS:  magnetic winding and 
the magnetorotational instability (MRI)~\cite{MRI0,MRI}.  Numerical simulations 
using a general relativistic magnetohydrodynamics
(GRMHD) code are required to follow this growth and determine the consequences.
The key subtlety is that the wavelength of the fastest-growing 
MRI mode must be well resolved on the computational grid.  Since this
wavelength is proportional to the magnetic field strength, it becomes
very difficult to resolve for small seed fields.  However, the 
simulations reported here succeed in resolving MRI.

New computational tools now make long-term numerical evolutions of 
relativistic magnetized HMNSs possible for the first time. Two groups 
have independently developed codes for evolving
MHD fluids in strong-field, dynamical spacetimes~\cite{DLSS2,SS} 
(see also~\cite{A05}). These codes solve the Einstein-Maxwell-MHD system 
of coupled equations, both in axisymmetry and in 3+1 dimensions, essentially
without approximation.  Both codes evolve the spacetime metric using the BSSN 
formulation~\cite{BSSN} and employ conservative shock-capturing
schemes to integrate the GRMHD equations.  Multiple tests have been performed
with these codes, including MHD shocks, MHD wave propagation, magnetized Bondi 
accretion, MHD waves induced by gravitational waves, and magnetized
accretion onto a neutron star.

To study the effects of magnetic fields on HMNSs, we first construct initial 
data assuming a $\Gamma=2$ polytropic EOS, $P=K\rho^{\Gamma}$, 
where $P$, $K$, and $\rho$ are the pressure, polytropic constant, and
rest-mass density. Henceforth, we adopt units such that $K=c=G=1$,
where $c$ is the speed of light and $G$ is the gravitational constant.
In these units, the maximum allowed baryon masses $M_0$ of nonrotating 
and of rigidly rotating stars are 0.180 and 0.207, respectively \cite{CST}.

Following previous papers (e.g, \cite{BSS,SBS,DLSS}), we choose the 
rotation law $u^t u_{\varphi}=A^2(\Omega_c-\Omega)$, where $u^{\mu}$ is 
the four-velocity, 
$\Omega \equiv u^{\varphi}/u^t$ is the angular velocity, and 
$\Omega_c$ is the angular velocity at the rotation axis.
The constant
$A$ has units of length and determines the degree of 
differential rotation. In this paper, $A$ is set equal to the
coordinate equatorial radius $R$, giving a value of $\sim 1/3$ for the
ratio of equatorial to central $\Omega$.

We model a
HMNS with $M_0=0.303$, maximum density $\rho_{\rm max}=0.0668$, 
and angular momentum parameter $J/M^2=1.00$~\cite{fn1}. The 
Arnowitt-Deser-Misner 
(ADM) mass is given by $M=0.279$, which is about 70\% 
larger than the OV limit. Because of 
rapid differential rotation, the shape of this star is highly flattened 
(see the first panel of Fig. 2).

During the evolution, we adopt a $\Gamma$-law (adiabatic) EOS 
$P=(\Gamma-1)\rho\varepsilon$ with $\Gamma=2$.  
Here, $\varepsilon$ denotes the specific internal energy. 
Before evolving, we add a weak poloidal magnetic field to the equilibrium model 
by introducing a vector potential of the form
$A_{\varphi}= \varpi^2 {\rm max}[A_b (P-P_{\rm cut}), 0]$, 
where $P_{\rm cut}$ is 4\% of the maximum pressure, $A_b$ 
is a constant that determines the initial strength of
the magnetic field, and $\varpi$ is the cylindrical radius. 
(A similar form of vector potential has been used in other 
MHD simulations~\cite{VP}.)
We characterize the strength of the initial magnetic field by 
$C\equiv {\rm max}(b^2/P)$, the maximum value on the grid of the 
ratio of the magnetic energy density $b^2$ to the pressure.  Several 
values of $A_b$ are chosen to yield the following values of $C$:
 $1.26 \times 10^{-3}$, $2.47 \times 10^{-3}$,
$4.88 \times 10^{-3}$, and $9.80 \times 10^{-3}$. We have verified 
that these small
initial magnetic fields introduce negligible violations of the 
Hamiltonian and momentum constraints. Note that $C \propto v_A^2$, 
where $v_A$ is the Alfv\'en speed. 
Thus, $C^{-1/2} \propto v_A^{-1}$ is proportional to the Alf\'ven time 
$t_A=R/v_A$. If the evolution time scale is determined by the
Alfv\'en time, a scaling relation in the evolution should hold.  
This relation may not hold for all $C$ due to MRI, which grows 
exponentially at a rate independent of $t_A$.
Comparing simulations with the values of $C$ quoted above, we indeed find that, 
if the time is rescaled as $C^{-1/2}t$, the results for different values 
of $C$ are approximately the same.  (Detailed results will be shown 
in~\cite{DLSSS1}.)  We therefore focus on the 
case with $C = 2.47 \times 10^{-3}$.  The typical value 
of $P$ for our model is $\sim 10^{34}~{\rm erg/cm^3}(2.8M_{\odot}/M)^2$;
the initial maximum magnetic field strength is then
$\sim 10^{16}(2.8M_{\odot}/M)$~gauss at $t=0$. (These scalings with $M$ 
assume our adopted initial polytropic model, for which $M/R = 0.22$).  
This magnetic field is too strong to model a typical HMNS (but is 
similar in strength to `magnetar' fields~\cite{MAGNETAR}).  
However, the qualitative behavior obtained here still applies
as long as the approximate scaling relation holds. 

\begin{figure}[thb]
\vspace{-4mm}
\begin{center}
\epsfxsize=3.in
\leavevmode
\epsffile{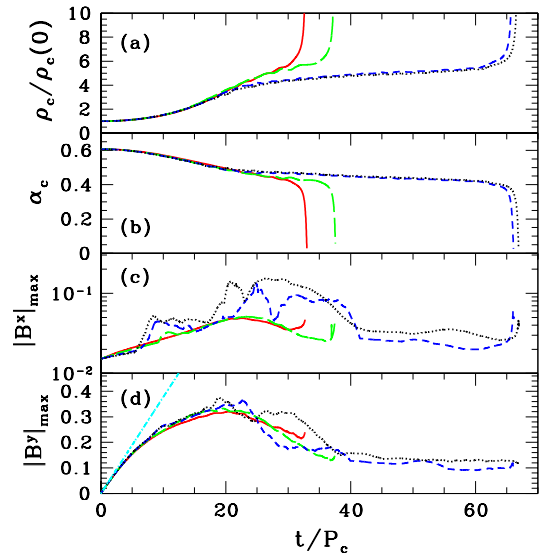}
\caption{Evolution of the central density, central lapse, and 
maximum values of $|B^x|$ and $|B^y|$ (the behavior of $|B^z|_{\rm max}$
is similar to the behavior of $|B^x|_{\rm max}$ and is therefore not shown). 
$|B^x|_{\rm max}$ and $|B^y|_{\rm max}$ are plotted in units of $\sqrt{\rho_{\max,0}}$
where $\rho_{\max,0}$ is the maximum rest-mass density at $t=0$. 
The solid (red), long-dashed (green), dashed (blue), and 
dotted (black) curves denote
the results with $N$=250, 300, 400, and 500 respectively.
The dot-dashed (cyan) line in the last panel represents the predicted
linear growth of $B^y$ at early times. 
\label{FIG1}}
\end{center}
\end{figure}

Simulations in axisymmetry were performed independently using two GRMHD 
codes~\cite{DLSS2,SS}, and the numerical results are
qualitatively similar. As in many 
hydrodynamic simulations, we add a tenuous 
``atmosphere'' to cover the computational grid outside the 
star. The atmospheric rest-mass density is set to 
$10^{-7} \rho_{\rm max,0}$ for the simulations shown here, 
where $\rho_{\rm max,0}$ 
is the maximum value of $\rho$ at $t=0$, which is 0.0668 in 
the adopted units.

We perform simulations on
a uniform grid with size $(N, N)$ in cylindrical
coordinates $(\varpi, z)$, which covers the region $[0,4.5R]$ for each
direction. For the HMNS adopted here, $R \approx
4.5M \approx 18.6~{\rm km}(M/2.8M_{\odot})$. 
To check the convergence of our numerical results, we perform 
simulations with four different grid resolution: $N$=250, 300, 400 and 500. 
We also checked that moving the outer boundary between $4R$ and $5R$
does not significantly affect the results.

Fig.~\ref{FIG1} shows the evolution of the central density $\rho_c$,
central lapse $\alpha_c$, and the maximum values of $|B^x|(\equiv
|B^{\varpi}|)$ and $|B^y|(\equiv \varpi |B^{\varphi}|)$ as functions of
$t/P_c$. Here $P_c \approx 39 M =0.54(M/2.8M_{\odot})$ ms 
denotes the central rotation period at $t=0$. The 
central density monotonically increases with time up to the 
formation of a black hole. Evolutions with various grid resolutions 
demonstrate that the results begin to converge when $N \agt 400$.
On the other hand, results are far from convergent for $N
\alt 300$. For example, the maximum values of $|B^x|$ are much smaller than
those with higher resolutions, and the growth rate of $|B^x|$ is
underestimated. Hence, the effect of MRI, 
which is responsible for the growth of $|B^x|$, is not computed 
accurately for low resolutions. This is because the wavelength of 
the fastest growing MRI mode is not well-resolved for low resolutions 
(see below).

\begin{figure*}[thb]
\begin{center}
\epsfxsize=1.8in
\leavevmode
\hspace{-0.7cm}\epsffile{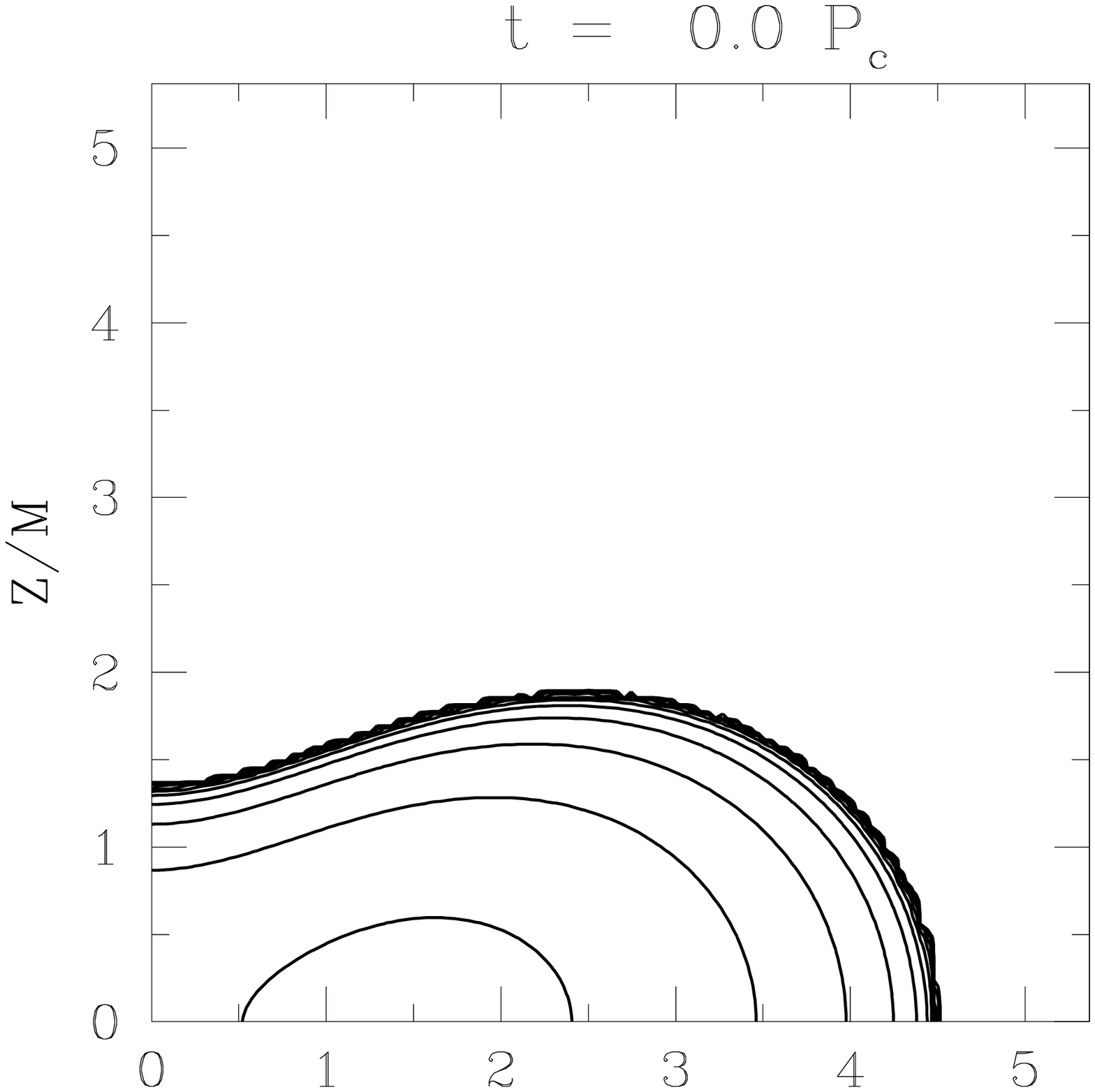}
\epsfxsize=1.8in
\leavevmode
\hspace{-0.7cm}\epsffile{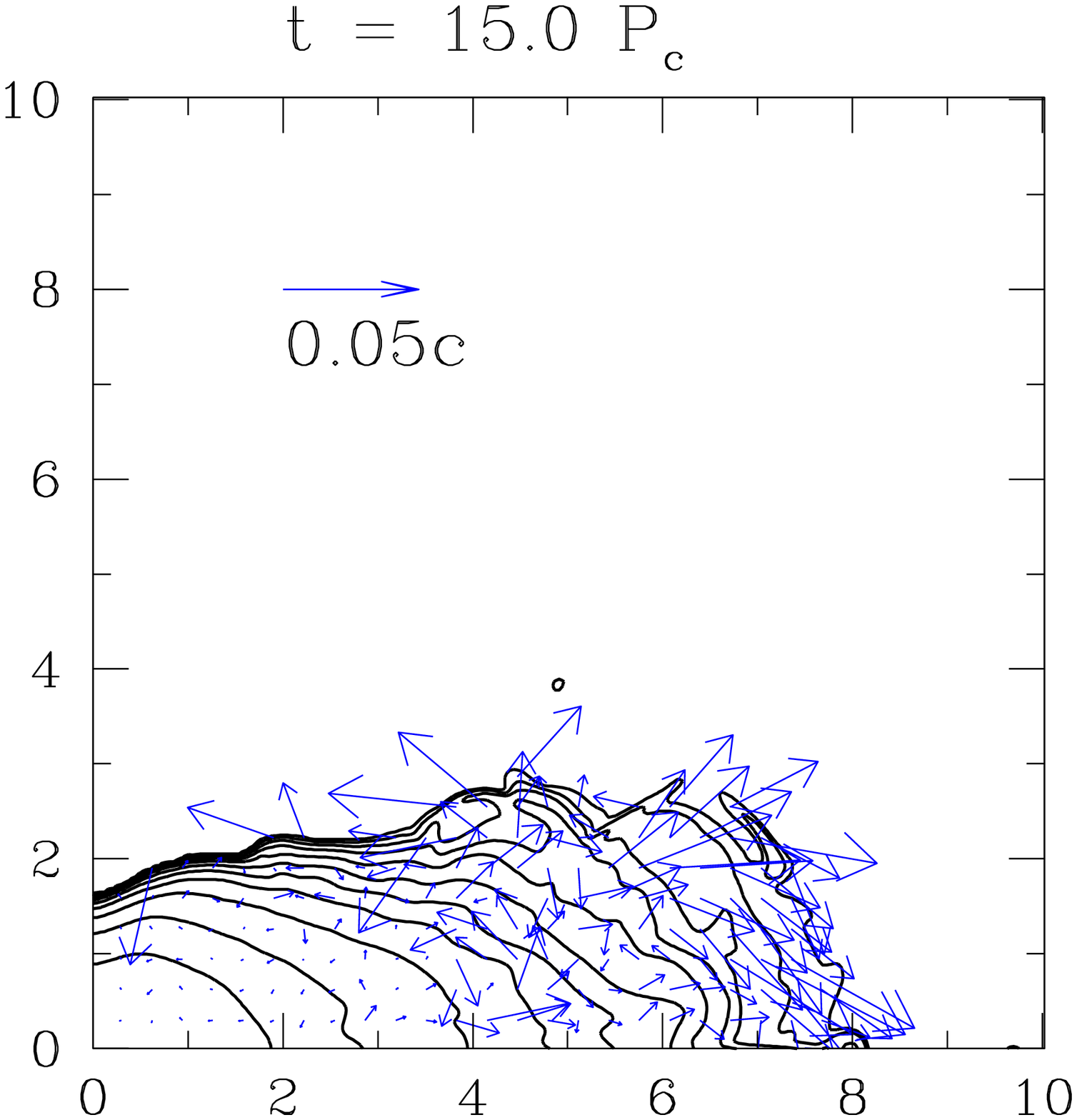}
\epsfxsize=1.8in
\leavevmode
\hspace{-0.7cm}\epsffile{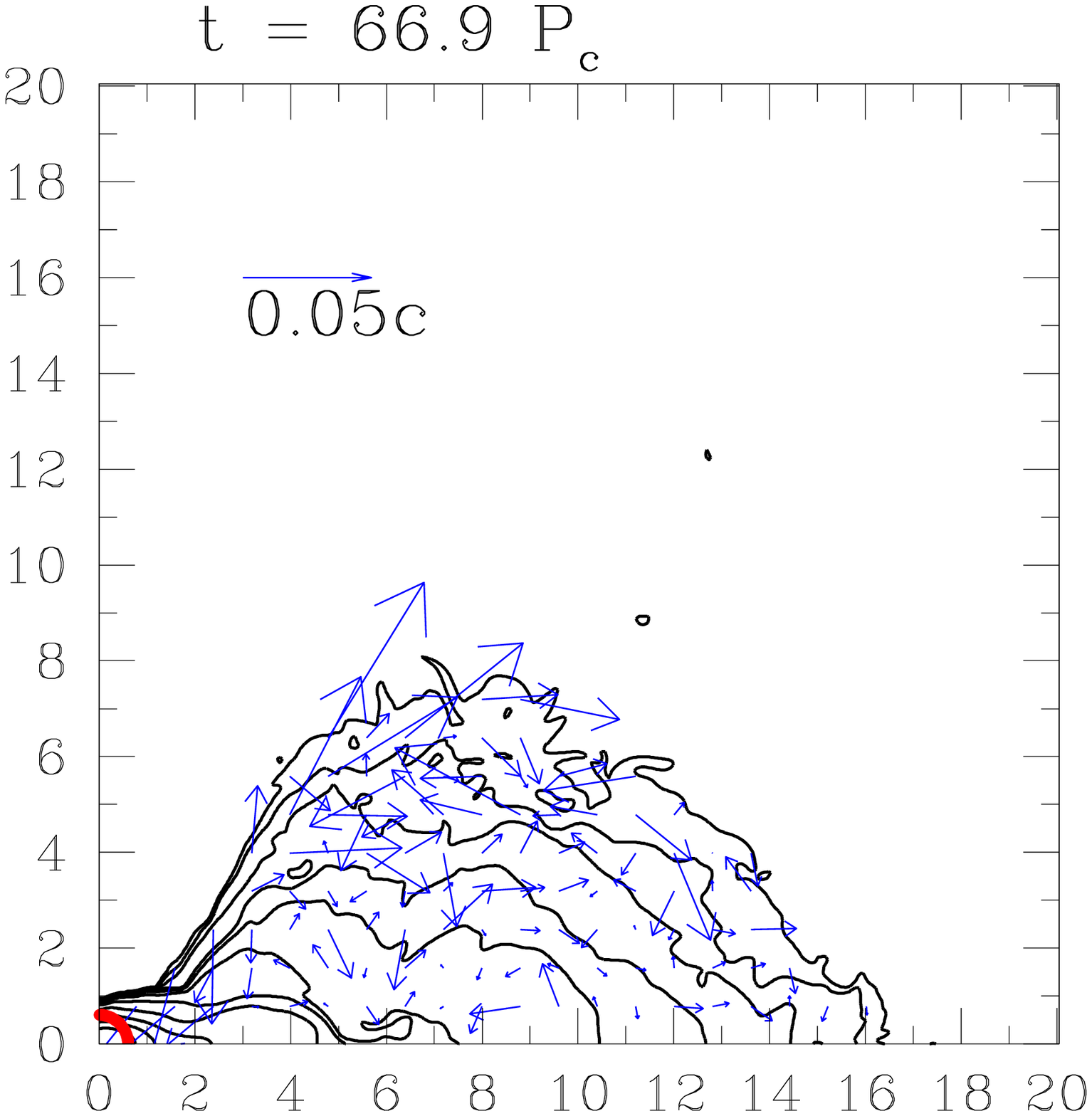}
\epsfxsize=1.8in
\leavevmode
\hspace{-0.7cm}\epsffile{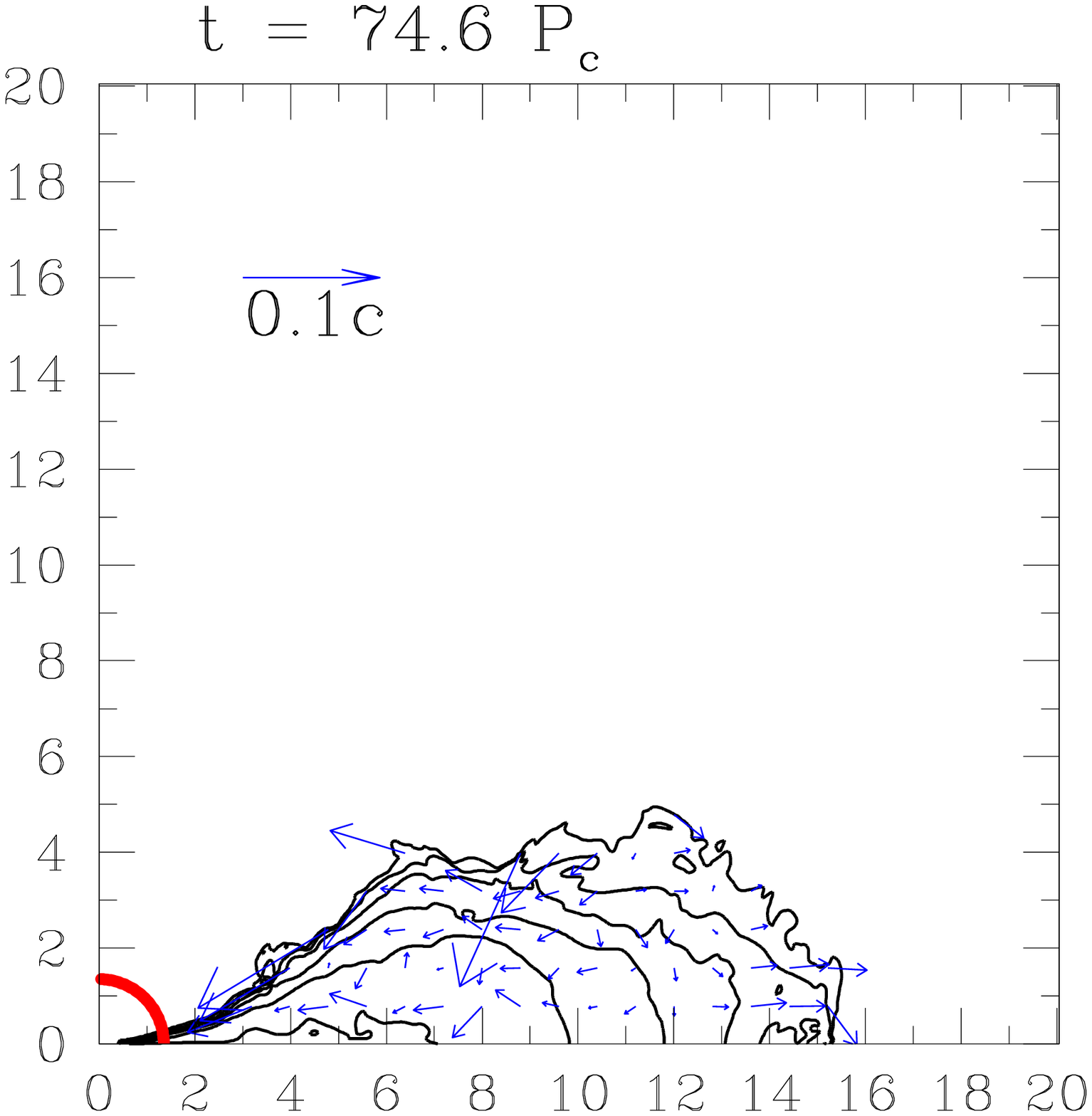} \\
\vspace{-0.5cm}
\epsfxsize=1.8in
\leavevmode
\hspace{-0.7cm}\epsffile{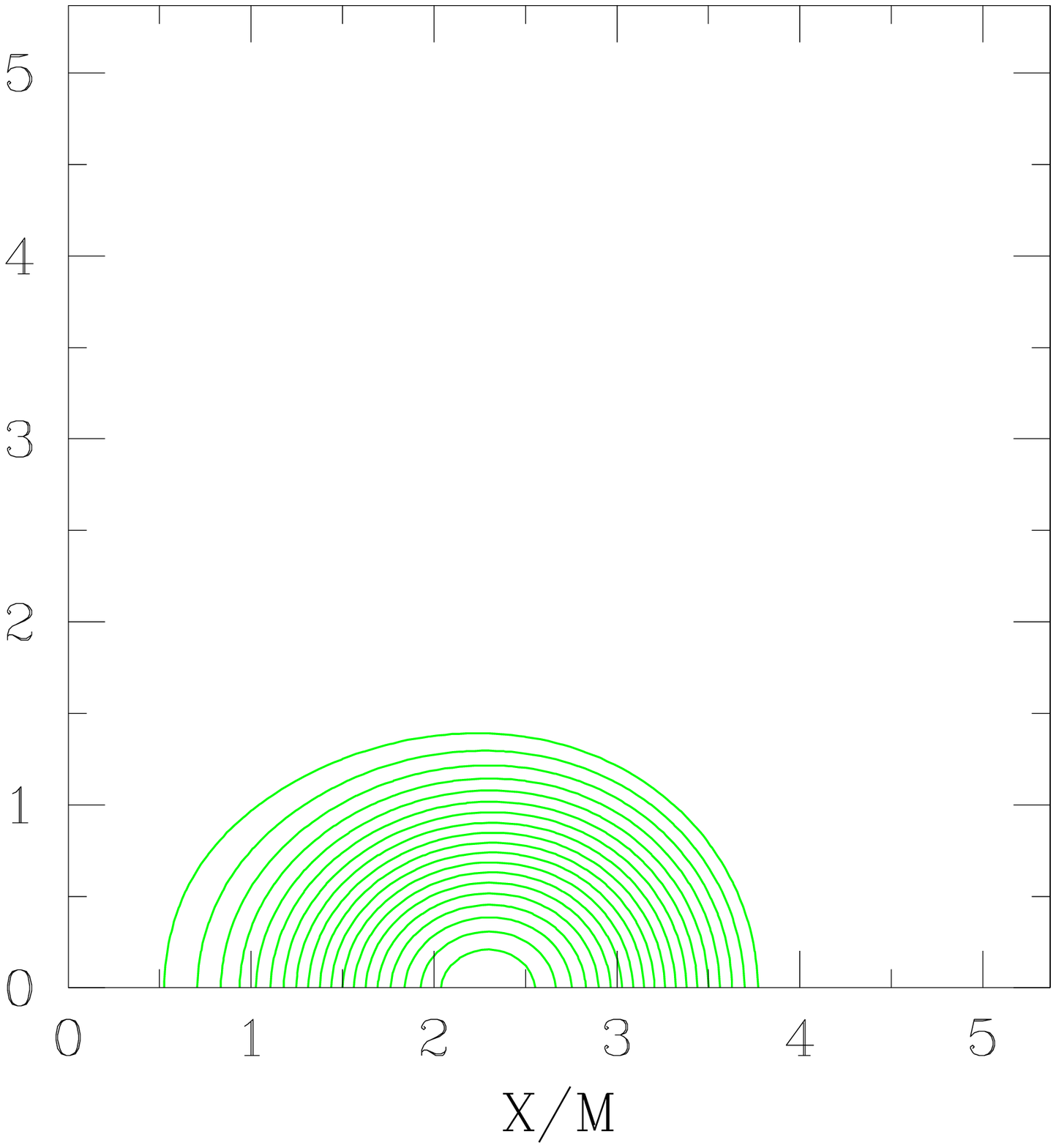}
\epsfxsize=1.8in
\leavevmode
\hspace{-0.7cm}\epsffile{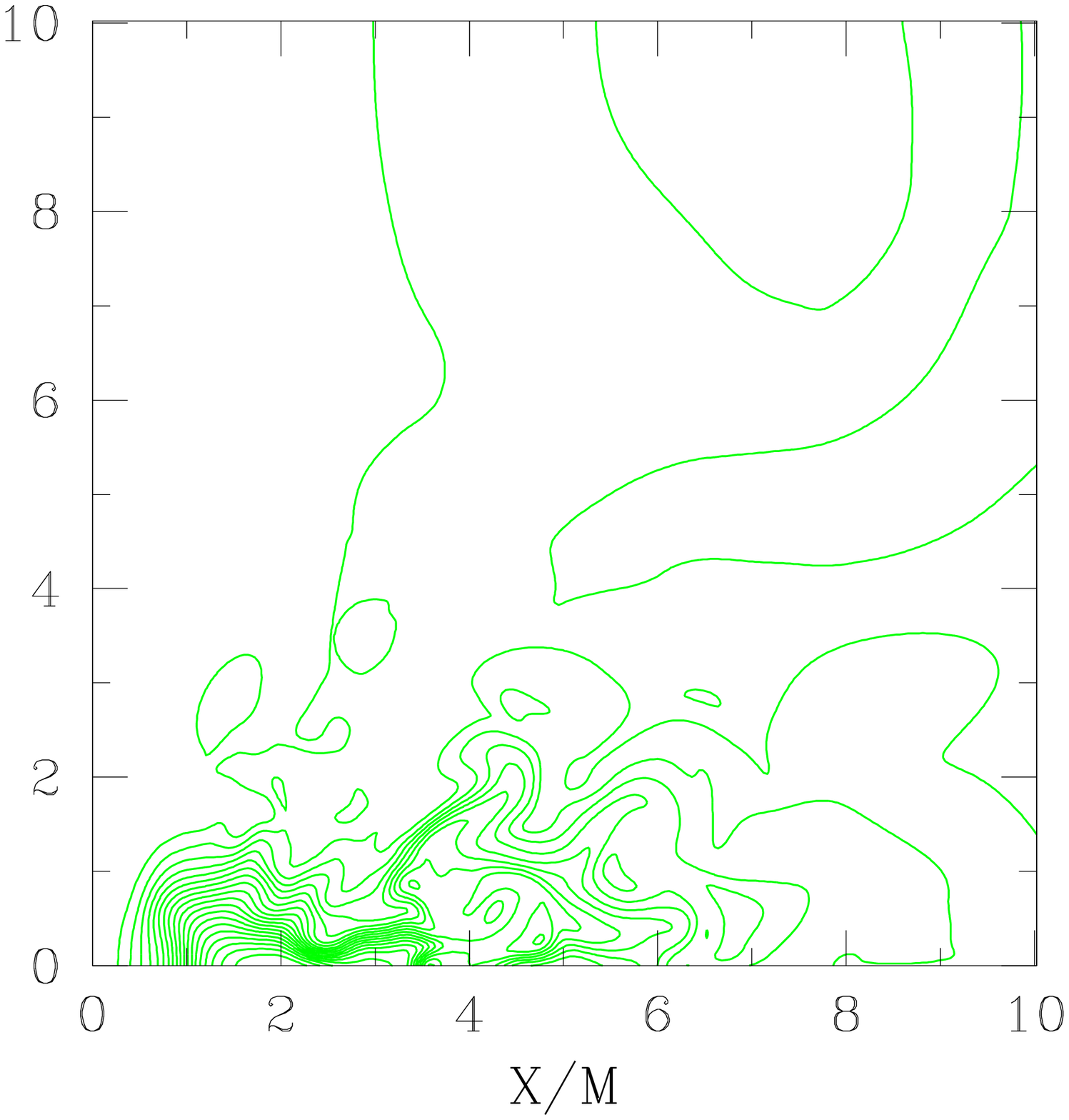}
\epsfxsize=1.8in
\leavevmode
\hspace{-0.7cm}\epsffile{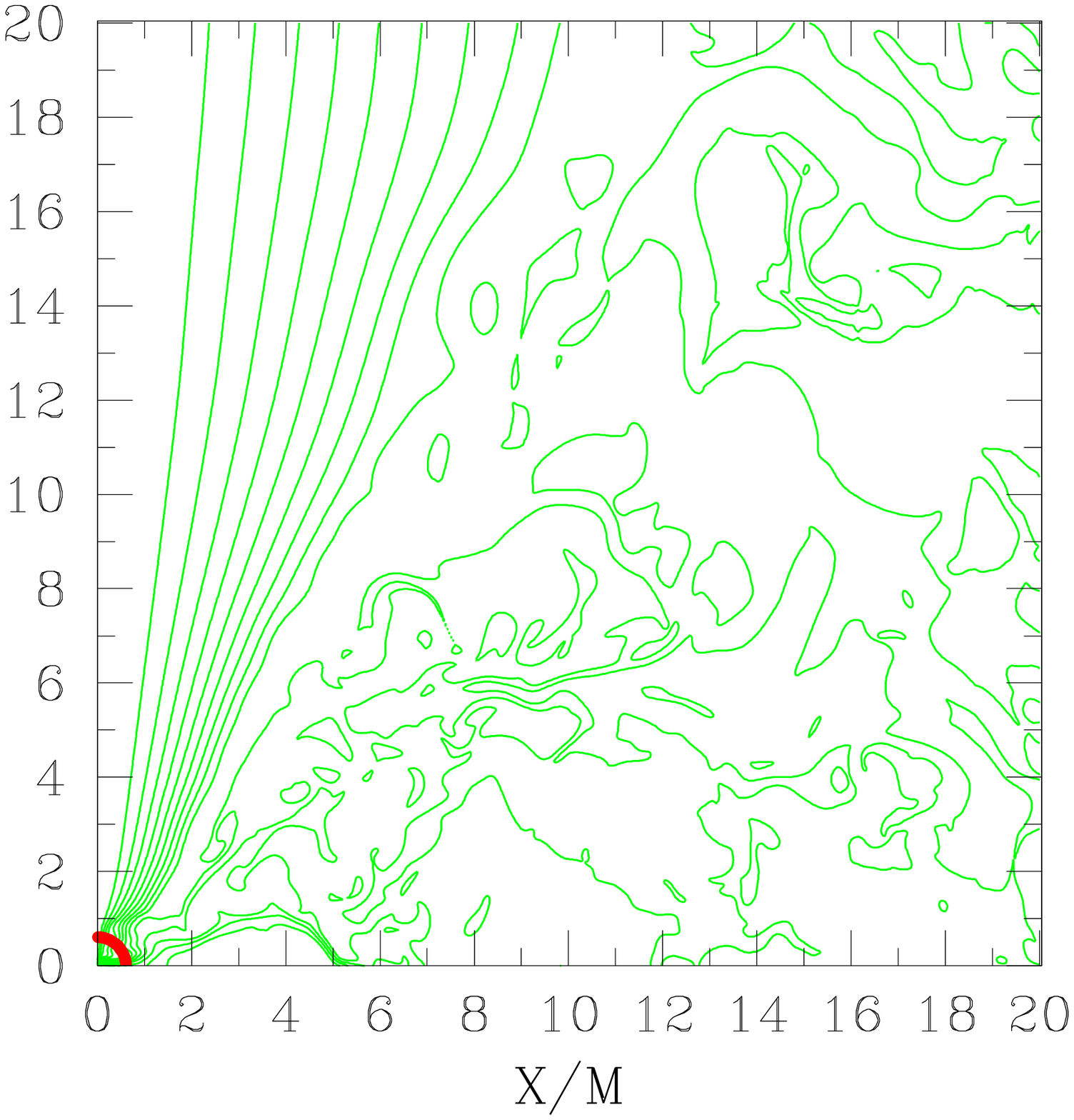}
\epsfxsize=1.8in
\leavevmode
\hspace{-0.7cm}\epsffile{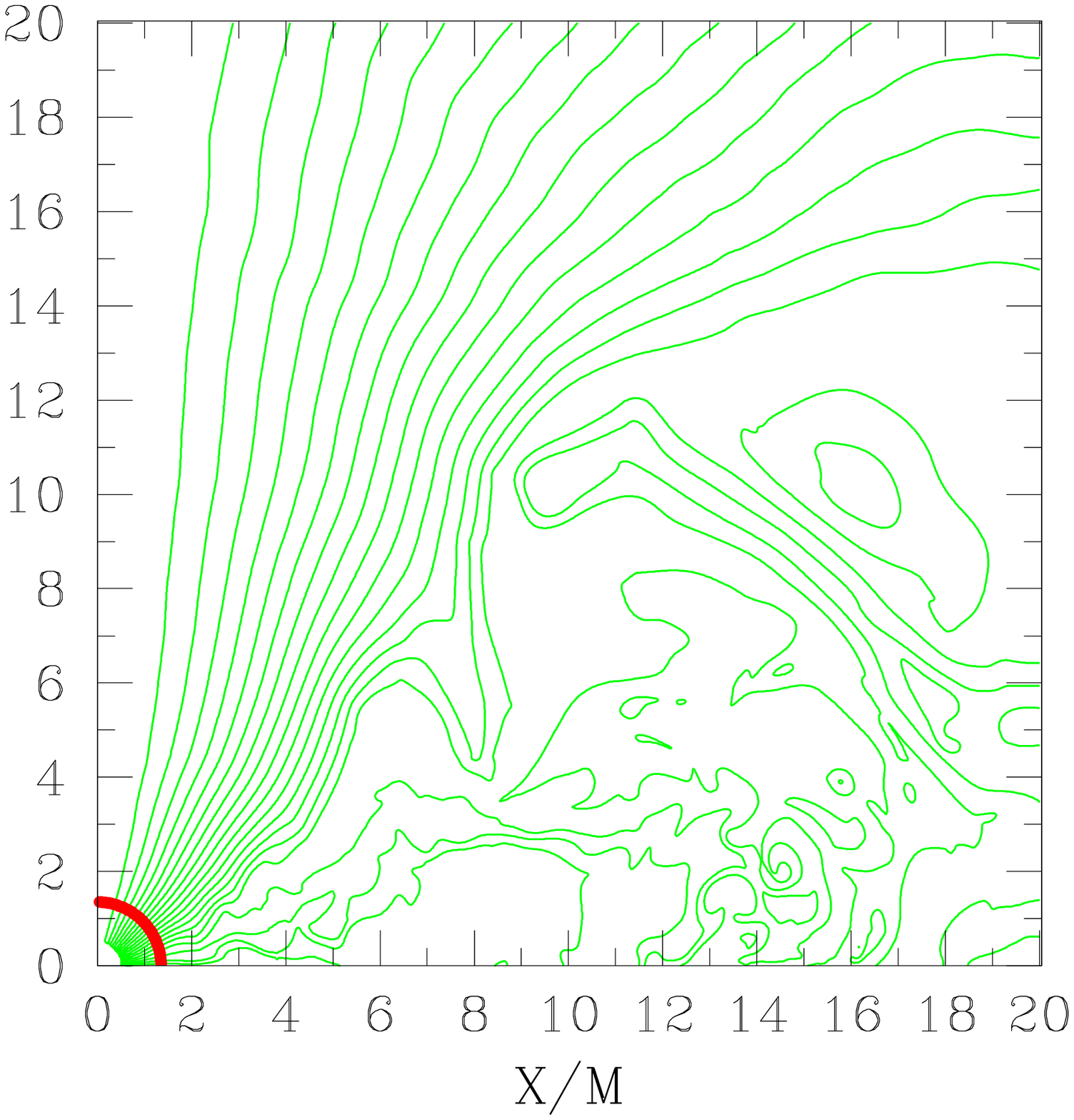}
\caption{The upper 4 panels show snapshots of the rest-mass density 
contours and velocity vectors on the meridional plane. The lower 
panels show the field lines (lines of constant $A_{\phi}$) 
for the poloidal magnetic field at the same times 
as the upper panels.  
The density contours are drawn for $\rho/\rho_{\rm max,0}=
10^{-0.3 i-0.09}~(i=0$--12).  
The field lines are drawn for $A_{\phi} = A_{\phi,\rm min} 
+ (A_{\phi,\rm max} - A_{\phi,\rm min}) i/20~(i=1$--19), 
where $A_{\phi,\rm max}$ and $A_{\phi,\rm min}$ are the maximum 
and minimum value of $A_{\phi}$ respectively at the given time. 
The thick solid (red) curves denote the apparent horizon.
\label{FIG3}}
\end{center}
\end{figure*}

Simulations with different treatments of the atmosphere 
are qualitatively the same when the atmospheric density is 
sufficiently small. However, the exact collapse time 
is somewhat sensitive to the details of numerical and atmospheric schemes. 
This is not surprising 
because, at late times, the star becomes marginally unstable, and  
the precise onset of collapse over the secular lifetime is sensitive 
to small differences in different schemes.

For the chosen initial strength of the seed magnetic field, 
the early evolution is dominated
by magnetic winding. When the seed field is weak, 
the induction equation shows that $B^y$ grows 
approximately linearly: $B^y(t;\varpi,z)\approx t \varpi B^i(0;\varpi,z) 
\partial_i \Omega(0;\varpi,z)$. Indeed, the early growth rate
agrees with the predicted one (cf.\ dot-dashed line in the last panel
of Fig.~{\ref{FIG1}). When the energy stored in the toroidal field becomes 
significant compared to the rotational energy, $|B^y|$ grows more slowly
and the degree of differential rotation is reduced. Eventually 
$|B^y|$ reaches a maximum and starts to decrease. This is expected to happen
on the Alfv\'en time scale $t_A$~\cite{Shapiro}, where the Alfv\'en
speed is $v_A =\sqrt{b^2/(\rho h+b^2)}$. For the model
considered here, the maximum value of $v_A=0.00734$, and thus, the
minimum value of $t_A=15.8 P_c$. We see that 
the maximum value of $|B^y|$ starts decreasing when $t \agt 20P_c$, 
consistent with the Alfv\'en time scale.

MRI is evident at times $t \alt 6P_c$ as shown in Fig.~\ref{FIG1}, 
where the maximum value of $|B^x|$ suddenly increases 
rapidly. MRI occurs wherever $\partial_{\varpi} \Omega < 0$~\cite{MRI}. 
The wavelength for the fastest growing mode is
$\lambda_{\rm MRI} \approx 2\pi v_A/\Omega$ and the e-folding time of 
the growth is 
$\tau_{\rm MRI} =2 \left(\partial \Omega/\partial 
\ln \varpi \right)^{-1}$~\cite{MRI}. With our choice of the initial
magnetic field strength, $\lambda_{\rm MRI}\sim R/10$ and 
$\tau_{\rm MRI} \sim 1P_c$. In Fig.~\ref{FIG1}, 
we see that MRI shows up prominently when $N \agt 400$. Hence, 
we need to use a resolution $\Delta \alt \lambda_{\rm MRI}/10$ 
to study the effect of MRI accurately. We find that 
MRI first occurs in the outer layers of the star near the equatorial 
plane. This can also be seen in Fig.~\ref{FIG3}, 
where we see that the magnetic field lines are distorted by $t=15P_c$.
In Fig.~\ref{FIG1}, 
we see that the the central density begins to grow more slowly once 
$|B^x|$ saturates. 
This may be caused by MRI-induced turbulence redistributing some of the 
angular momentum to slow down the contraction of the core.

The combined effects of magnetic braking due to winding
and MRI eventually trigger 
gravitational collapse to a black hole at
$t \approx 66P_c \approx 36 (M/2.8M_{\odot})$ ms when an apparent 
horizon forms. 
The latest simulations~\cite{STUb} of BNS mergers show 
that for a sufficiently stiff EOS and typical observed 
BNS masses, HMNS formation is possible.
HMNS remnants are triaxial and strong emitters of
gravitational waves in these simulations. The dissipation time scale 
of angular momentum due to
gravitational radiation is $\sim 100$~ms~\cite{STUb}.
Therefore, HMNSs with an initially large magnetic field 
($B \agt 10^{16}$G) will be subject to  `delayed' collapse due to MHD 
effects (magnetic braking + MRI) rather than
by the emission of gravitational waves. For seed magnetic fields which are much
weaker than the cases studied here,
gravitational radiation may be the trigger of collapse. 
However, it is possible that MRI may dominate the 
evolution even in this case, since the e-folding time of MRI is independent
of the initial field strength. A more careful study of this scenario has to be
carried out in full 3D~\cite{fn2}. However, since any dissipative agent 
(viscosity, magnetic fields,
gravitational radiation) serves to redistribute and/or carry off angular
momentum, the final fate of an HMNS -- collapse to a black hole, 
accompanied by a gravitational wave burst -- is assured.

Soon after the formation of the apparent horizon, the simulations 
become inaccurate because of grid stretching. 
To follow the subsequent evolution, 
a simple excision technique for black hole spacetimes is 
employed~\cite{excision,DLSS0}.  The evolution of the
irreducible mass of the black hole computed from the area of the
apparent horizon $A_{\rm AH}$ as 
$M_{\rm irr}=\sqrt{A_{\rm AH}/16\pi}$, and the total
rest mass outside the apparent horizon are shown in 
Fig.~\ref{FIG4}.  Soon after 
formation, the black hole grows rapidly, swallowing the surrounding
matter. However, the accretion rate $\dot M_0$ gradually decreases and
the black hole settles down to a quasiequilibrium state, i.e., the
growth time scale becomes much longer than the dynamical
time scale. At the end of the simulation, $\dot M_0$ decreases
to $\approx 0.01 M_0/P_c$. The estimated value of the black hole 
spin parameter is $J_{\rm hole}/M_{\rm hole}^2 \sim 0.8$.
The black hole angular momentum 
is computed from $J_{\rm hole}=J-J_{\rm matter}(r>r_H)$, where 
the angular momentum of the matter outside the horizon $J_{\rm matter}(r>r_H)$ 
is computed by a volume integral (see e.g., Eq.~(51) of~\cite{DLSS}). 
The mass of the black hole $M_{\rm hole}$ is crudely estimated from   
$M_{\rm hole}\approx \sqrt{M_{\rm irr}^2 + (J_{\rm hole}/2M_{\rm irr})^2}$. 
The density contour curves and magnetic field lines 
at the end of the 
simulation are shown in the last column of panels of Fig.~\ref{FIG3}. 

The value of $\dot{M}_0$ indicates that the accretion time scale is $\sim
10$--$20P_c \approx 5$--$10~{\rm ms} (M/2.8M_{\odot})$.  Also, we find 
that the specific internal thermal energy in the torus 
near the surface is substantial because of shock
heating, indicating that the torus can be a strong emitter of
neutrinos. These facts suggest that the system 
formed after the `delayed' collapse of a magnetized
HMNS (black hole + hot torus + collimated magnetic field) 
is a candidate for the central engine of
short gamma-ray bursts~\cite{GRB1,AJM}.  This possibility is explored in 
more detail in~\cite{GRB2}.

\begin{figure}[thb]
\vspace{-4mm}
\begin{center}
\epsfxsize=2in
\leavevmode
\epsffile{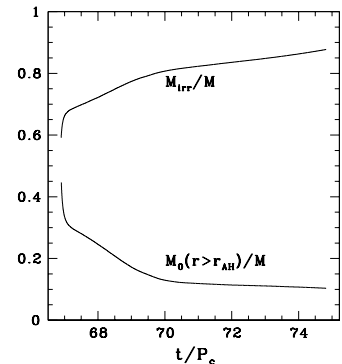}
\vspace{-5mm}
\caption{Evolution of the irreducible mass 
and the total rest-mass outside the apparent horizon.
\label{FIG4}}
\end{center}
\end{figure}

\acknowledgments
It is a pleasure to thank C.~Gammie for useful suggestions and discussions.
Numerical computations were performed at the National Center for 
Supercomputing Applications at the University 
of Illinois at Urbana-Champaign (UIUC), and 
on the FACOM VPP5000 machine at
the data analysis center of NAOJ and the NEC SX6 machine in ISAS, 
JAXA. This work was in part supported by NSF Grants PHY-0205155 
and PHY-0345151, NASA Grants NNG04GK54G and NNG046N90H 
at UIUC, and  
Japanese Monbukagakusho Grants (Nos.\ 17030004 and 17540232).


\begin{thebibliography}{99}

\bibitem{BSS} 
T. W. Baumgarte, S. L. Shapiro, and M. Shibata, 
Astrophys. J. Lett. {\bf 528}, L29 (2000).

\bibitem{RS99} F. A. Rasio and S. L. Shapiro, Astrophys.\ J.\  
{\bf 432}, 242 (1994); Class. Quant. Grav. {\bf 16} R1 (1999).

\bibitem{FR00} J. A. Faber and F. A. Rasio, Phys. Rev. D {\bf 62}, 
064012 (2000).

\bibitem{STUa} M. Shibata, K. Taniguchi, and K. Ury\=u,
Phys. Rev. D {\bf 68}, 084020 (2003).

\bibitem{STUb} M. Shibata, K. Taniguchi, and K. Ury\=u, 
Phys. Rev. D {\bf 71}, 084021 (2005).

\bibitem{DLSS} M. D. Duez, Y. T. Liu, S. L. Shapiro, and B. C. Stephens,
Phys. Rev. D {\bf 69}, 104030 (2004). 

\bibitem{DLSS2} M. D. Duez, Y. T. Liu, S. L. Shapiro, and B. C. Stephens,
Phys. Rev. D {\bf 72}, 024028 (2005). 

\bibitem{SS} M. Shibata and Y.-I. Sekiguchi, 
Phys. Rev. D {\bf 72}, 044014 (2005). 

\bibitem{A05} L. Ant\'on, O. Zanotti, J. A. Miralles, J. M. Mart\'i,
J. M. Ib\'a\~nez, J. A. Font, and J. A. Pons, Astrophys. J., in press,
astro-ph/0506063 (2005).

\bibitem{BSSN} M. Shibata and T. Nakamura, Phys. Rev. D {\bf 52}, 5428 
(1995); T. W. Baumgarte and S. L. Shapiro, Phys. Rev. D {\bf 59}, 
024007 (1999).

\bibitem{CST} G. C. Cook, S. L. Shapiro, and S. A. Teukolsky, 
Astrophys. J. {\bf 422}, 227 (1994).

\bibitem{SBS} M. Shibata, T. W. Baumgarte, and S. L. Shapiro,
Astrophys. J. {\bf 542}, 453 (2000).  

\bibitem{fn1} The star chosen here is
the same as that referred to as star~I in~\cite{DLSS}.

\bibitem{VP} J-P. De Villiers, J. F. Hawley, and J. H. Krolik, 
Astrophys.\ J., {\bf 599}, 1238 (2003); 
J. C. McKinney and C. F. Gammie, Astrophys.\ J., {\bf 611}, 
977 (2004) 

\bibitem{DLSSS1} M. D. Duez, Y. T. Liu, S. L. Shapiro, M. Shibata,
and B. C. Stephens, in preparation. 

\bibitem{MAGNETAR} R. C. Duncan and C. Thompson, Astrophys.\ J.\ 
Lett., {\bf 392}, 9 (1992).

\bibitem{Shapiro} S. L. Shapiro, Astrophys. J. {\bf 544}, 397 (2000);
J. N. Cook, S. L. Shapiro, and B. C. Stephens, Astrophys. J. {\bf 599},
1272 (2003).

\bibitem{MRI0} V. P. Velikhov, Soc. Phys. JETP, {\bf 36}, 995 (1959); 
S. Chandrasekhar, Proc.\ Natl.\ Acad.\ Sci.\ USA, {\bf 46}, 253 (1960).

\bibitem{MRI} S. A. Balbus and J. F. Hawley, Astrophys. J. {\bf 376},
214 (1991); Rev. Mod. Phys. {\bf 70}, 1 (1998). 

\bibitem{fn2} The growth of MRI may be significantly enhanced when 
axisymmetry is relaxed; see 
J. F. Hawley, Astrophys. J. {\bf 528}, 462 (2000).

\bibitem{excision}
M. Alcubierre and B. Br\"ugmann, Phys. Rev. D {\bf 63}, 104006 (2001). 

\bibitem{DLSS0} M. D. Duez, S. L. Shapiro, and H-J.\ Yo,
Phys. Rev. D {\bf 69}, 104016 (2004). 

\bibitem{GRB1} E.g., T. Piran, Phys. Rep. {\bf 314}, 575 (1999); 
{\em ibid} {\bf 333}, 529 (2000): Rev. Mod. Phys. {\bf 76}, 1143 (2005).  

\bibitem{AJM} M. A. Aloy, H.-T. Janka, and E. M\"uller, Astron. Astrophys. 
{\bf 436}, 273 (2005). 

\bibitem{GRB2} M. Shibata, M. D. Duez, Y. T. Liu, S. L. Shapiro, and
B. C. Stephens, submitted to Phys.\ Rev.\ Lett.\ (astro-ph/0511142).

\end{thebibliography}
\end{document}